\documentclass[5p]{elsarticle}
\usepackage{amssymb}
 \usepackage{amsthm} 
 \usepackage{amsmath}
\usepackage{graphicx,color}
\usepackage{bm}

\journal{Solid State Communications}
\begin{document}
\begin{frontmatter}
\title{Superconducting proximity effect in interacting quantum dots revealed by shot noise}

\author{Alessandro Braggio\corref{cor1}}
\ead{alessandro.braggio@spin.cnr.it}
\address{CNR-SPIN, Dipartimento di Fisica, Universit\`a di Genova, Genova, Italy}

\author{Michele Governale}
\address{School of Chemical and Physical Sciences and MacDiarmid Institute for Advanced Materials and Nanotechnology, Victoria 
University of Wellington, Wellington, New Zealand}

\author{Marco G. Pala}
\address{IMEP-LAHC (UMR 5130), Grenoble INP, Minatec, Grenoble, France}

\author{J\"urgen K\"onig}
\address{Theoretische Physik, Universit\"at Duisburg-Essen and CeNIDE, Duisburg, Germany}

\begin{abstract}
We study the full counting statistics of charge transport through a quantum dot tunnel-coupled to one normal and one 
superconducting lead with a large superconducting gap.
As function of the level detuning, there is a crossover from a regime with strong superconducting correlations in the 
quantum dot to a 
regime in which the proximity effect on the quantum dot is suppressed.
We analyze the current fluctuations of this crossover in the shot-noise regime.
In particular, we predict that the full counting statistics changes from Poissonian with charge $2e$,  typical for Cooper pairs, to 
Poissonian with charge $e$, when the superconducting proximity effect is present. Thus, the onset of the superconducting 
proximity effect is revealed by the reduction of the Fano factor from 2 to 1.
\end{abstract}



\end{frontmatter}

\section{Introduction}
Current fluctuations in a nanoscale conductor yield more information than the average current  on the  processes involved in 
transport and on the nature of the current-carrying excitations. For this reason the study of shot  noise has attracted a lot of 
attention both theoretically and experimentally \cite{blanter:2000,nazarovB:2003}. 
The full current distribution function is 
given in terms of the Full Counting Statistics (FCS), $P(N,t)$, defined as the probability that $N$ charges have passed through the 
system in a given time $t$ \cite{levitov:1993,levitov:1996}.
The FCS can be conveniently expressed in terms of the Cumulant Generating Function (CGF), 
\begin{equation}
S(\chi)=\ln\left[\sum_{N=-\infty}^\infty \exp (i N \chi) P(N,t)\right] \, ,
\end{equation}
from which the current cumulants are obtained by performing derivatives with respect to the counting field $\chi$, i.e. $
\kappa^{(n)}= (e^n /t) 
\left. \partial_{i\chi}^n S(\chi)\right|_{\chi=0}$, with $e<0$ 
being
the electron charge. Several theoretical approaches have been 
developed to determine the CGF both 
in non-interacting 
and
in interacting nano-scale systems \cite{nazarov:2002,
nagaev:2002,belzig:2001,nazarov:2003,pilgram:2003,jordan:2004,bagrets:2003,braggio:2006,emary:2007,schaller:2007,flindt:2008,flindt:2010}.
The measurement of high-order moments of the current is a highly-demanding experimental task 
\cite{reulet:2003,lindell:2004,bomze:2005,huard:2007,timofeev:2007,gershon:2008}.
A breakthrough has been recently achieved by 
measuring in real time single-electron tunneling through a quantum dot by means of a nearby quantum point contact 
\cite{gustavsson:2006,fricke:2007,fujisawa:2006,flindt:2009,gustavsson:2009}.
The FCS is an important tool to study transport in hybrid normal-superconducting systems 
\cite{belzigB:2003,borlin:2002,choi:2001,choi:2003,cuevas:2003,johansson:2003},
since in this case it becomes of fundamental importance to identify whether the correlations induced by the pairing in the 
superconductor persist in the normal regions.
 
Hybrid superconducting-normal structures with quantum dots offer the possibility to study the interplay of superconducting 
correlations, strong Coulomb interaction, quantum fluctuations and non-equilibrium physics as a function of tunable system 
parameters.
Subgap transport in these structures is mediated by Andreev 
reflection \cite{degennes:1963,andreev:1964} 
and it has been studied extensively both 
theoretically \cite{fazio:1998,fazio:1999,kang:1998,schwab:1998,clerk:2000,rozhkov:2000,tanaka:2007,karrasch:2008,meng:2009,jonckheere:2009, 
zazunov:2010,cuevas:2001,avishai:2001,golubev:2007,golubev:2009,pala:2007,governale:2007,futterer:2009} and 
experimentally \cite{morpurgo:1999,morpurgo:2002,cleuziou:2002,herrero:2002,jorgensen:2006,vandam:2006,jespersen:2007,buizert:2007,jorgensen:2007,eichler:2009,hofstetter:2009,herrmann:2010,deacon:2010,deacon:2010a}.

In this letter, we study the FCS of electrons transferred through a single-level quantum dot tunnel coupled to one normal and one 
superconducting lead with a large superconducting gap. 
The generalized Fano factor $F_j=\kappa^{(j)}/(-e)^{j-1}\kappa^{(1)}$ for Andreev reflection at a normal-super\-conductor 
interface is $F_j=2^{j-1}$, indicating 
the transfer of Andreev pairs~\cite{muzykantskii:1994}.
As we will discuss below, the Fano factors remain $F_j=2^{j-1}$ for many transport regimes when a quantum dot is placed in 
between.
For strong tunnel coupling between dot and superconductor, however, the Fano factor can be 
reduced to $F_j=1$. 
In that case, indeed, the proximity effect leads to fast coherent oscillations of Cooper pairs between dot and superconductor, 
which are 
stochastically interrupted by injection of single electrons from the normal lead. 
Once the proximity effect is suppressed either by a large energy detuning of the empty and double-occupied dot, or by a large 
tunnel coupling to the normal lead, the Fano factors goes back to $F_j=2^{j-1}$. 
This counter-intuitive result shows that the superconducting proximity effect on the dot can be revealed by a measurement of 
current fluctuations, rather than the current.

\section{Theory}
We consider the Hamiltonian
$H=H_{\rm dot}+\sum_{\eta=N,S}(H_\eta+H_{\rm tunn,\eta}) $, where $\eta=N,S$ labels the leads.  
The dot is described by the Anderson impurity model with a spin-degenerate level $\varepsilon$ and on-site Coulomb repulsion $U$
\begin{equation} 
H_{\rm dot}= \varepsilon \sum_\sigma d_{\sigma}^\dagger d_\sigma +U
n_{\uparrow}n_{\downarrow},  
\end{equation}
where $d_{\sigma} (d_{\sigma}^\dagger)$ is the annihilation(creation) operator for an electron with spin
$\sigma=\uparrow,\downarrow$ and $n_{\sigma}=d^{\dagger}_{\sigma}d_{\sigma}$. 
The dot eigenstates are denoted as $|0\rangle$ (empty), $|\sigma\rangle=d_\sigma^\dagger|0\rangle$ (singly occupied with spin $\sigma$), and 
$|D\rangle= d_\uparrow^\dagger d_\downarrow^\dagger |0\rangle$ (doubly occupied). 
The Hamiltonian of lead $\eta=N,S$ reads
\begin{equation}
H_{\eta}=\!\sum_{k \sigma} \epsilon_{\eta k}
c_{\eta k \sigma}^\dagger c_{\eta k \sigma}- \delta_{\eta,S}\Delta\! \sum_{k}\left(  c_{\eta -k \downarrow} 
c_{\eta k \uparrow}+\textrm{H.c.}\right), 
\end{equation}
with the lead-electron operators $c_{\eta k \sigma}$ and $c_{\eta k \sigma}^\dagger$, and 
$\delta_{\eta,S}=1$ for $\eta=S$ and $0$ for $\eta=N$ . 
Superconductivity in lead $S$ is described by means of the mean-field BCS Hamiltonian with a pair potential 
$\Delta = g \sum_k \langle c^\dagger_{\eta k \uparrow}c^\dagger_{\eta -k \downarrow}\rangle $, where $g$ denotes
 the interaction strength. 
Without loss of generality, $\Delta$ has been taken real and positive. 
We choose the electrochemical potential of the superconductor as a reference $\mu_S=0$. The normal lead has 
electrochemical potential $\mu_N=\mu$.
The tunnel coupling between dot and leads is modeled by the tunneling Hamiltonians: 
\begin{equation}H_{\rm tunn,\eta}=V_{\eta}\sum_{k \sigma}c^{\dagger}_{\eta
  k\sigma}d_{\sigma}+\mathrm{H.c.}
  \end{equation} 
We assume that the normal-state density of states $\nu_\eta$ of lead $\eta$ is constant in the energy window relevant for transport and  define the tunnel-coupling strength as $\Gamma_\eta=2\pi \nu_\eta |V_\eta|^2$.
Since we are interested in sub-gap transport, we consider the limit $\Delta\rightarrow \infty$, in which the superconducting lead's 
degrees of freedom can be integrated out.
This results in the effective Hamiltonian that for the dot plus the superconducting lead \cite{rozhkov:2000}
\begin{equation}
\label{Heff}
H_{\rm eff}= H_{\rm dot}- \frac{\Gamma_{S}}{2} \left( d^\dagger_{\uparrow} d^\dagger_{\downarrow}  + d_{\downarrow} 
d_{\uparrow}\right) .
\end{equation}
The tunnel coupling to the superconductor mixes the states $|0\rangle$ and $|D\rangle$ to form new eigenstates $|\pm\rangle= 
\frac{1}{\sqrt{2}}\sqrt{1\mp\frac{\delta}{2\epsilon_A}} |0\rangle \mp\frac{1}{\sqrt{2}}\sqrt{1\pm\frac{\delta}{2\epsilon_A}} |D\rangle$ 
with energies $\delta/2\pm \epsilon_A$.
Here $\delta=2\varepsilon+U$ is the detuning between empty and doubly-occupied state 
and $2\epsilon_A=\sqrt{\delta^2+ \Gamma_S^2}$ the splitting between the $|+\rangle$ and the $|-\rangle$ states.
The excitation energies of the dot define the Andreev addition energies 
\begin{equation}
\label{eq:AndreecBStates}
E_{A, \gamma', \gamma} = \gamma' U/2 + \gamma \epsilon_A ,
\end{equation}
with $\gamma', \gamma = \pm 1$, whose position can be identified by measuring the charge or the current as a function of the 
electrochemical potential $\mu$ \cite{pala:2007}, as long as $\Gamma_N \ll k_B T\ll \epsilon_A$. 

\section{Calculation}
We trace out the normal lead's degrees of freedom in order to obtain an effective description of the dot in terms of its reduced 
density matrix.  
For $\Gamma_N \ll \epsilon_A$, the reduced density matrix is diagonal in the eigenbasis of the effective Hamiltonian (\ref{Heff}).
The vector of the diagonal matrix elements, the occupation probabilities, at time $t$ under the condition that $N$ electrons have 
passed through the barrier between the normal lead and the dot, $\mathbf{p}(N,t)$, obeys the generalized master equation 
\cite{braggio:2006}
\begin{equation} 
\frac{d}{dt} \mathbf{p}(N,t)=\!\!\sum_{N'=-\infty}^\infty\! \int_0^t dt' \mathbf{W}(N-N',t-t') \mathbf{p}(N',t') ,
\end{equation}
where the Kernel  $\mathbf{W}(N-N',t-t')$ describes transitions involving the transfer of $N-N'$ electrons. 
The counting field is introduced through the Fourier transform $\mathbf{W}^{\chi}(t-t')=\sum_N  \mathbf{W}(N,t-t') \exp(i \chi N)$. 
In the paper we consider only the weak-coupling regime $\Gamma_N\ll k_B T$ and we perform a perturbation expansion to first 
order in $\Gamma_N$. To this order of perturbation theory non-Markovian corrections play no role {\cite{braggio:2006} and we need 
only the zero-frequency Laplace transform of the Kernel:   $\mathbf{W}^{\chi}=\int_0^\infty  \mathbf{W}^{\chi}(\tau) d\tau$. 
The long-time behavior is determined by the eigenvalue of $\mathbf{W}^{\chi}$ with the smallest  absolute value of the real part,  
$\lambda(\chi)$ \cite{bagrets:2003,braggio:2006}.
In this regime the CGF in first order in $\Gamma_N$ is  given by  $S=t \lambda^{(1)}(\chi)$, where $t$ is the measurement time 
and  the superscript explicitly indicates first order in $\Gamma_N$. 
A general diagrammatic theory for computing the Kernel $\mathbf{W}^{\chi}$ exists  
\cite{pala:2007}, 
but in the case under consideration the rates $W_{\psi\psi'}^{\chi}$ for transitions from the dot state $\psi'$ to the state $\psi$ can be 
computed straightforwardly  by means of Fermi golden rule introducing a factor $\exp(-i\chi)$ $[\exp(i\chi)]$ for each electron 
leaving [entering] the normal lead. 
The rates read: 
\begin{eqnarray}
W_{\sigma,\pm}^{\chi}=\frac{\Gamma_N}{2}\!\!\sum_{\gamma=\pm1}\!\!\left( 1\mp\gamma \frac{\delta}{2\epsilon_A}\right) f^{\gamma}(E_{A,-\gamma,\mp\gamma})\  \mathrm{e}^{ -i\gamma\chi}\\
W_{\pm,\sigma}^{\chi}=\frac{\Gamma_N}{2}\!\!\sum_{\gamma=\pm1}\!\!\left( 1\pm\gamma \frac{\delta}{2\epsilon_A}\right) f^{\gamma}(E_{A,\gamma,\pm\gamma})\  \mathrm{e}^{ -i\gamma\chi},
\end{eqnarray}
where $f^{+}(\omega)$ is the Fermi function of the normal lead with electrochemical potential $\mu$, and 
$f^{-}(\omega)=1-f^{+}(\omega)$. The diagonal elements of the Kernel do not contain the counting field and are given by 
$W_{+,+}=-2W_{\sigma,+}^{\chi}|_{\chi=0}$, $W_{-,-}=-2W_{\sigma,-}^{\chi}|_{\chi=0}$, and 
$W_{\sigma,\sigma}=-(W_{+,\sigma}^{\chi}+W_{-,\sigma}^{\chi})|_{\chi=0}$.

\section{Results}
\begin{figure}
\begin{center}
\includegraphics[width=.99\columnwidth,clip=true]{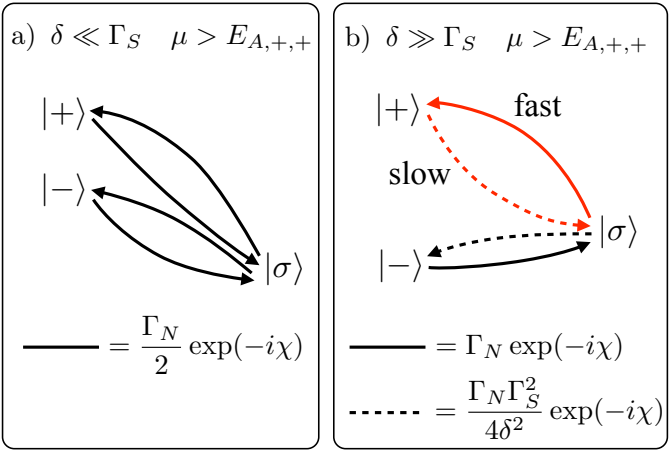}
\end{center}
\caption{Schematic description of the rates contributing to transport in the high-bias regime. 
a) On resonance, all rates are the same and all give rise to the transfer of one electron from the normal lead to the dot. The states 
$|\pm\rangle$ are coherent superposition of $|0\rangle$ and $|D\rangle$ with equal weight: 
$|\pm\rangle=\frac{1}{\sqrt{2}}\left(|0\rangle\mp |D\rangle\right)$. 
b) Off resonance there are fast processes (solid lines) and slow processes (dashed lines). For $\delta\gg \Gamma_S$, the dot will 
be mostly stuck in the state $|+\rangle\approx
\frac{\Gamma_S}{2\delta}|0\rangle+|D\rangle$. Transport occurs along the red cycle (lighter gray) and it consists of  uncorrelated 
tunneling of pairs of electrons. 
}\label{rates}
\end{figure}
We discuss the regime $|\delta|<\sqrt{U^2-\Gamma_S^2}$ for which the Andreev bound states obey the inequality 
$E_{A,+,+}>E_{A,+,-}>E_{A,-,+}>E_{A,-,-}$. In the following we consider only the 
low-temperature limit $k_B T\ll\epsilon_A$ and 
the case of positive $\mu$. The results for negative $\mu$ can be obtained by performing the substitutions $\delta \rightarrow -
\delta$ and $\chi \rightarrow-\chi$.
In the high-bias regime, $\mu > E_{A,+,+}$, 
the CGF reads
\begin{equation}
	S_{\rm hb}=t \Gamma_N \Biggl[\sqrt{\frac{\delta^2+\Gamma_S^2 \mathrm{e}^{ -2i\chi} }{\delta^2+\Gamma_S^2}} -1 \Biggr].
\end{equation}
On resonance, $|\delta|\ll \Gamma_S$, this expression simplifies  to the CGF corresponding to Poissonian tunneling  with 
strength $\Gamma_N$ of electrons through a barrier
\begin{equation}
S_{\rm hb,on}= t \Gamma_N \left( \exp(-i\chi)-1 \right) \, ,
\end{equation}
while off resonance, $|\delta|\gg \Gamma_S$, we obtain
\begin{equation}
S_{\rm hb,off}= t \Gamma_N \frac{\Gamma_S^2}{ 2\delta^2}\left( \exp( -2i \chi) -1\right), 
\end{equation}
which corresponds to a Poissonian transfer of Andreev pairs. 
In between the two limits, the FCS is non-Poissonian. We now proceed to clarify the physical picture that leads to this somewhat 
counter-intuitive result.\\ 
On resonance ($|\delta|\ll\Gamma_N$): In the large-bias regime, the tunneling rates are all equal and they are given by 
\begin{equation}
W_{\sigma,\pm}^{\chi}=W_{\pm,\sigma}^{\chi}\approx(\Gamma_N /2)\exp(-i\chi). 
\end{equation}
The situation is pictorially shown in panel a) of figure~\ref{rates}.
The stationary probability of being in the $|+\rangle $, $|-\rangle $ and $|\sigma\rangle$  states are the same and equal to $1/4$. 
Suppose that at a given time the dot is in the singly-occupied state $|\sigma\rangle$.
On average, after an interval of time of $2\hbar/\Gamma_N$ an electron from the normal lead enters the dot.
The resulting doubly-occupied state, $|D\rangle = (|-\rangle - |+\rangle)/\sqrt{2}$, undergoes fast coherent oscillations to 
$|0\rangle$ and back, describing tunneling of an electron pair forth and back between quantum dot and superconductor.
This happens well before than another electron enters 
in the dot after a time $2\hbar/\Gamma_N$. 
Seen from the normal conductor, single electrons hop on the dot independently with a rate $\Gamma_N/2$.\\ 
\begin{figure}
\begin{center}
\includegraphics[width=.99\columnwidth,clip=true]{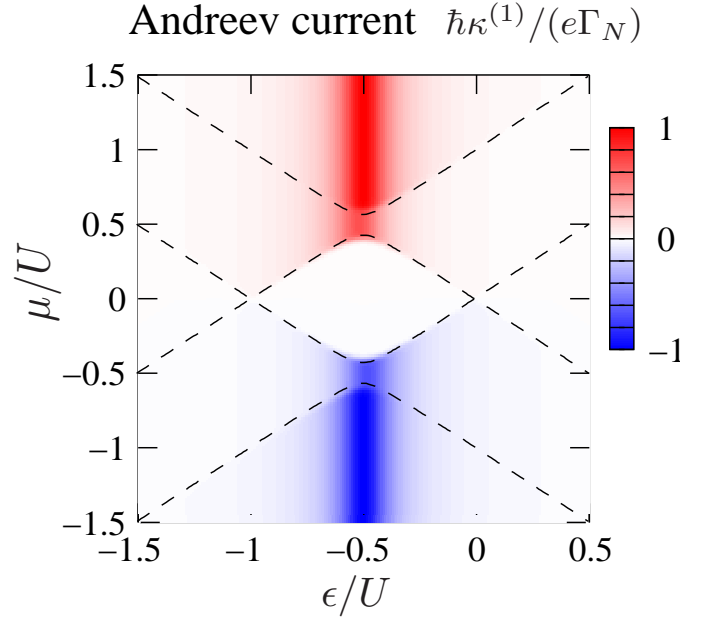}\\ 
\end{center}
\caption{
Current in the normal lead in units of $e \Gamma_N/\hbar $ as a function of level position $\epsilon$ and of chemical potential 
$\mu$ for $\Gamma_S=0.2\, U$, $k_B T=0.01\, U$ and $\Gamma_N \ll \Gamma_S$. The dashed lines map the Andreev 
bounds states as $\mu=E_{A, \gamma', \gamma}$.}
\label{figcurrent}
\end{figure}
Off resonance ($|\delta|\gg\Gamma_N$): For simplicity we discuss only the $\delta>0$ case.
The rates are given by
 \begin{eqnarray*}
W_{\sigma,+}^{\chi}=W_{-,\sigma}^{\chi}\approx &\Gamma_N \Gamma_S^2/(4 \delta^2)\exp(-i\chi)\\
W_{+,\sigma}^{\chi}=W_{\sigma,-}^{\chi}\approx &\Gamma_N\exp(-i\chi)\ ,
\end{eqnarray*}
and the dot will be with a high probability in the state 
$|+\rangle$. 
Panel b) of figure~\ref{rates} elucidates the situation. 
Notice that in this regime 
the state $|+\rangle$ corresponds mostly 
to $|D\rangle$  with only a small admixture of $|0\rangle$. 
The dot first needs to inject a Cooper pair in the superconductor via a 
cotunneling process (factor $\Gamma_S^2/\delta^2$) and only then an electron can tunnel in the dot from the normal side (factor $
\Gamma_N$.)
This combined process is the bottleneck for transport and leads to a suppression of the rate $W_{\sigma,+}^{\chi}$, which goes as $\Gamma_N \Gamma_S^2/\delta^2$. 
On the contrary, when the dot is in $|\sigma\rangle$, the rate 
for tunneling back to $|+\rangle$ is much higher $\propto \Gamma_N$ and the dot will return quickly to the state $|+\rangle$. 
Seen from the normal side, pairs of electrons are injected with an average time separation of 
$4\hbar\delta^2/(\Gamma_N \Gamma_S^2)$. The average separation of the electrons within a pair is much shorter, namely
$\hbar/\Gamma_N$.

In the intermediate-bias regime, $E_{A,+,-}<\mu<E_{A,+,+}$, the situation is more 
complex.
The CGF reads
\begin{equation}
\label{intermediate}
S_{\rm ib}=t \frac{\Gamma_N}{8\epsilon_A} \biggl[-\delta-6\epsilon_A+\sqrt{(3\delta+2\epsilon_A)^2+8\Gamma_S^2 
\mathrm{e}^{ -2i \chi}}\biggr].
\end{equation}
The Poissonian one-electron transport regime is now realized for $\delta=-\frac{\sqrt{2}}{4}\Gamma_S$ where the CGF becomes 
$S_{\rm ib} =  t\frac{ 2 \Gamma_N}{3} \left( \exp( -i\chi)-1 \right)$. Off resonance, we recover the Poissonian tunneling of Andreev pairs
\begin{equation}
S_{\rm ib,off}= t \Gamma_N 
\frac{\Gamma_S^2}{ \delta^2}
\frac{1}{3+\mathrm{sgn} \, \delta}
\left( \exp( -2i \chi) -1\right).
\end{equation}
This behavior can be understood looking at the rates in a similar fashion as for the high-bias case.
Finally, in the low-bias regime, $0< \mu < E_{A,+,-}$, transport is suppressed by Coulomb blockade.
In figure~\ref{figcurrent} we show the current that, as already discussed in 
reference~\cite{pala:2007}, exhibit a dependence on the Andreev 
addition energies (dashed lines) as a function of $\epsilon$ and $\mu$. 

In figure~\ref{figcumulants}, for fixed bias, we show the transition from Fano factors $F_j\approx2^{j-1}$ to $F_j\approx1$ for 
$F_2$(black) and $F_3$(red) as a function of $\epsilon$. Off resonance, we have $F_j\approx2^{j-1}$, 
indicating that electrons are transfered as Andreev pairs.
On resonance, the superconducting proximity effect leads to fast coherent oscillations of a Cooper pair between dot 
and superconductor. 
Electrons are injected from the normal lead via uncorrelated single-electron tunneling events, 
$F_j\approx1$.
In the intermediate-bias regime (dot-dashed lines in figure~\ref{figcumulants}) we see a similar behavior but with a shifted position 
of the minimum.

\begin{figure}
\begin{center}
\includegraphics[width=.99\columnwidth,clip=true]{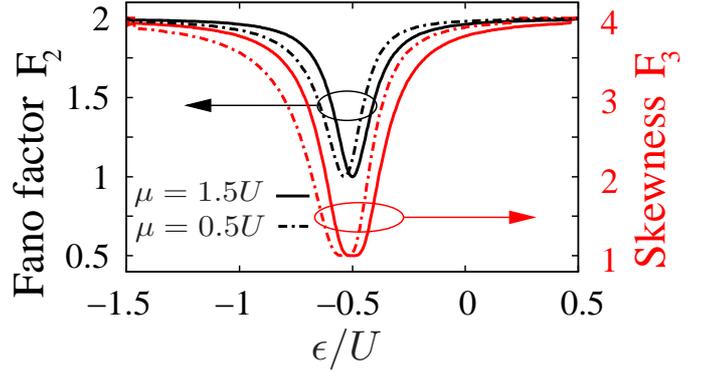}\\ 
\end{center}
\caption{
Fano factor $F_2$ (black lines) and normalized skewness $F_3$ (red lines) as a function of level position for $\mu=1.5\, U$ 
(solid line) and $\mu=0.5\, U$ (dot-dashed line).  
The other parameters are the same as in figure~\ref{figcurrent}.}
\label{figcumulants}
\end{figure}

In the analysis presented above no dephasing giving rise to a suppression of the superconducting proximity effect is included.
In fact, in the limit of weak tunnel coupling to the normal lead, $\Gamma_N \ll \epsilon_A$, as assumed above, the Andreev bound states are sharp resonances without broadening.
The regime of a quantum dot more strongly coupled to a superconductor than to a normal lead, i.e. $\Gamma_S \gg \Gamma_N$,  has already been realised experimentally by Deacon \textit{et al.} \cite{deacon:2010,deacon:2010a}.
The effect of dephasing can be studied by increasing the tunnel coupling to the normal lead. 
In the extreme limit of $\Gamma_N \gg \epsilon_A$, the superconducting proximity effect is fully suppressed,
and the picture of uncorrelated electron-pair tunneling remains valid for all values of $\delta/\Gamma_S$.
We explicitly checked the cases $\Gamma_N \gg |\delta| \gg \Gamma_S$ and $\Gamma_N \gg \Gamma_S \gg |\delta| $ in the 
high-bias regime (by means of the diagrammatic theory \cite{pala:2007} and including matrix elements of $\mathbf{W}$ that are 
off-diagonal in the $|\pm\rangle$ basis) to find the CGF 
\begin{equation}
S =  t\frac{ \Gamma_S^2}{2\Gamma_N} \left( \exp(-2i\chi)-1 \right)\ , 
\end{equation}
that corresponds again to Poissonian transfer of Andreev pairs with the rate $ \Gamma_S^2/2\Gamma_N$. 
For intermediate values of $\Gamma_N$, representing finite but not complete dephasing, we expect an intermediate value of the Fano factor between 1 and 2 on resonance. In that sence, the measurement of the Fano factor may serve as a tool to determine the presence and strength of dephasing.

\section{Conclusions}
We have computed the CGF for transport through a quantum dot coupled to one normal and one 
superconducting lead in the large-gap limit. In the high-bias and intermediate bias regimes, there is a crossover from generalized 
Fano factors $F_j=2^{j-1}$ to $F_j=1$, which clearly identifies the superconducting proximity effect in the quantum dot.

\section{Acknowledgement}
 We acknowledge financial support from the
DFG via SFB 491, and from INFM-CNR via Seed Project PLASE001.

\end{document}